\documentclass[12pt,aps]{revtex4}

\input{epsf}
\begin{document}

\author{G.G.Kozlov}

\title{The algorithm for simulating of phase transition in Ising magnetic}
\begin{abstract}
Simple algorithm of dynamics of Ising magnetic is described.
The algorithm can be implemented on conventional  digital
computer and can be used for construction of specialized
processor for simulation of ferromagnetic systems. The
algorithm gives a simple way to calculate 1D correlation
functions for 1D Ising magnetic.
\end{abstract}
\maketitle
\section{Introduction}

In recent past the analog computers were the only tool for
solving the mathematical problems which could not be solved analytically.
 The analog computer is a device (typically - an electric circuit)
whose temporal  dynamics is described by  equations similar to
those which are to be solved. The analog computers have the virtue
of being simple. The main defect of analog computers is
absence of universality.
For this reason for solving of any particular problem one must design  a
special analog computer suited for solving only this particular problem.
 The impressive achievements of semiconductor
technology of recent decades made it possible  to create a digital
universal computer based on microprocessor - a programmable
electronic device which is millions times more complicated than
any of analog computers but allows one to solve wide range of
problems just by entering of appropriate program.  Despite the
great power of modern digital universal microprocessors one can
point out some problems whose solution require excessive time or
even impossible. Having in mind one of these problems one can put
the following question:  is it possible (using the
fantastic facilities of up-to-date technology) to create a
specialized device (specialized processor) suited for solving only
this particular problem? Imagine that we have constructed the
specialized processor for solving of many-particle problem of
atomic physics. In our opinion despite the loss of universality
this processor would be of  great  interest.   Below we suggest
some algorithm for solving the problem of phase transition in
Ising magnetic. This algorithm can be directly implemented by means of
conventional digital computer. On the other hand in our opinion it is quite
possible to create the specialized processor working according  to this
algorithm but much faster than conventional universal processor.
This processor being much more simple device than the
universal processor allows one  to solve the problem of phase
transition in Ising system and to obtain an arbitrary values of
interest (energy, magnetisation, heat capacity). We estimate the
complexity of this processor to be comparable with that  of modern
memory devices.

Let us consider a trigger -- scheme with two stable states which we
denote by $\pm 1$. This trigger and spin $1/2$ have much in common
and below we will not differ these two.
  Suppose this trigger can change its state only when clock pulse
  coming. We  consider the train of clock pulses  to be
  equidistant in time.  Consider the probability for trigger
  to change its state (the probability of corresponding spin flip)
  be depending on the trigger's state before clock pulse coming.
 Consequently the probability $p_+$ for trigger to switch
from  state $s=+1$ differs from that $p_-$ to
 switch from  state $s=-1$:
$p_+\ne p_-$.
 Let these probabilities obey:
 \begin{equation}
 p_\pm=\ae \exp \bigg(\beta h s\bigg), \hskip10mm s=\pm 1.
 \end{equation}

Equation (1) has the sense of detail equilibrium principle
for  spin $1/2$ in magnetic field $h$ and
in contact with thermostat with inverse temperature $\beta$.
 Since  $p_\pm\le 1$ the normalization constant $\ae$ must obey:

\begin{equation}
\ae\le\exp \bigg( -\beta |h| \bigg)
\end{equation}

If we now consider the statistical ensemble of such
triggers then the kinetic equations for average
number $n_\pm$ of triggers in states $\pm 1$ have the form:

 \begin{equation}
\Delta n_+^i=-n_+^i p_+ + n_-^i p_-
 \end{equation}
$$
\Delta n_-^i=-n_-^i p_- + n_+^i p_+
$$

where $\Delta n^i_\pm$ -- is  increment of number of triggers
in  state $s=\pm 1$ after $i$-th clock pulse coming,
$n^i_ \pm$ -- number of triggers in state $s=\pm 1$ to
 the moment of $i$-th clock pulse coming.

The steady state solution of this equation has the form:
\begin{equation}
{n^{st}_+\over n^{st}_-}={p_-\over p_+}=\exp \bigg(-2\beta h \bigg)
\end{equation}

 This corresponds to thermal
 equilibrium state of spin system in the external magnetic field $h$.

Now let us consider the case of Ising magnetic i.e. the
 lattice comprised of
$N$ spins (triggers) coupled to each other in such a
  way that the magnetic
 field acting on the arbitrary  spin is defined by
  configuration of the rest spins in the lattice.
In the simplest case the magnetic field $h_i$ acting on  spin with
number $i$  is produced by its nearest neighbours and if we denote
 the set of nearest neighbours by $nn(i)$ then:

\begin{equation}
h_i=W\sum_{r\in nn(i)} s_r,
\end{equation}
here the constant $W$  characterize  interspin coupling.
 By the analogy with the aforesaid let us consider the following
dynamics of this system.
 The clock pulses act sequentially  on all
the spins (triggers) in the lattice -- we call this {\it round trip}.
 During the  round trip any spin (say $i$-th) may be  overturned  with
 probability defined by equation (1) with magnetic
field $h_i$ defined by formula (5).
 We are interesting in the dynamics of this system under the action of the
 train of round trips.
 For the above dynamics of this "Ising magnetic" (consisting of
triggers controlled by clock pulses)  we now will obtain
the kinetic equation for  density matrix and will show
that its steady state solution corresponds to  thermal
equilibrium. By the analogy with real magnetic let us describe the state
 of our "magnetic" (consisting of triggers) by the wavefunction
  whose $i$-th argument describe the state of $i$-th spin (trigger):

\begin{equation}
\Psi=|s_1,s_2,...s_N\rangle
\end{equation}

Introduce $\hat{O}_i$ -- operator of $i$-th spin flip:
\begin{equation}
\hat{O}_i |s_1,s_2,...,s_i,...,s_N\rangle=|s_1,s_2,...,-s_i,...,s_N\rangle
\end{equation}

Let us introduce the statistical ensemble of Ising magnetics
and let  $\sigma(\Psi)$ be the number of magnetics in  state $\Psi$ in this ensemble.
 Up to normalization factor the quantities $\sigma(\Psi)$ are represent the
 diagonal elements of the density matrix of Ising magnetic.
 Let us consider $\Delta\sigma^i$ -- the increment of $\sigma$ when
 clock pulse act on $i$-th spin (trigger)

\begin{equation}
\Delta\sigma^i(\Psi)=-\sum_{\Phi\ne\Psi} \sigma(\Psi)V_{\Psi\rightarrow\Phi}
+\sum_{\Phi\ne\Psi} \sigma(\Phi)V_{\Phi\rightarrow\Psi}
\end{equation}

 Here $V_{\Psi\rightarrow\Phi}$ -- is the probability of  transition
   from $\Psi$-state to $\Phi$-state when clock pulse act on $i$-th
 spin.
 In accordance with the above dynamics of Ising
 magnetic the only non-zero probabilities are:

\begin{equation}
(\hbox{probability of  transition from $\Psi$ to $\Phi=\hat{O}_i\Psi$})=\ae\exp \bigg(\beta
h_i(\Psi)s_i(\Psi)\bigg)
\end{equation}
\begin{equation}
(\hbox{probability of  transition from $\Phi=\hat{O}_i\Psi$ to $\Psi$})=\ae\exp \bigg(\beta
h_i(\Phi)s_i(\Phi)\bigg)=
\end{equation}

$$
=\ae\exp \bigg(-\beta h_i(\Psi)s_i(\Psi)\bigg)
$$
 The last equality  follows from the fact that the field
acting on $i$-th spin in  states $\Phi=\hat{O}_i\Psi$ and $\Psi$
 is the same while
the value of $i$-th spin has the opposite sign
(i.e. $s_i(\Phi=\hat{O}_i\Psi)=-s_i(\Psi)$).
 Then using  equation (8) one can see that

 \begin{equation}
 \Delta\sigma^i(\Psi)=\ae\bigg[
 \sigma(\hat{O}_i\Psi)\exp\bigg(-\beta h_i(\Psi)s_i(\Psi) \bigg)-
\sigma(\Psi)\exp\bigg(\beta h_i(\Psi)s_i(\Psi) \bigg)
 \bigg]
 \end{equation}

Let us show that this equation has the steady state solution in the form:

\begin{equation}
\sigma_{eq}(\Psi)=\exp\bigg(\lambda H(\Psi)\bigg)
\end{equation}
where $H$ -- is the Hamiltonian of Ising magnetic:
\begin{equation}
H={W\over 2}\sum_{i=1}^N\sum_{\alpha=nn(i)}s_i s_\alpha
\end{equation}
 We need to calculate $\sigma_{eq}(\Phi=\hat{O}_i\Psi)$. To do this note that:
 \begin{equation}
 H(\Phi=\hat{O}_i\Psi)=H(\Psi)-2W
  \sum_{\alpha=nn(i)}s_i s_\alpha\bigg |_\Psi=
H(\Psi)-2h_i(\Psi)s_i(\Psi)
 \end{equation}

 Now calculating  $\sigma_{eq}(\Phi=\hat{O}_i\Psi)$ by equation (12) and
substituting the result in to equation (11) it is easy to see that when
$\lambda=-\beta$ the right part of equation (11)  vanishes. So we see that (12) is the
steady state solution of (11) and represent  thermal
equilibrium density matrix of Ising magnetic.
 Thus the above algorithm of sequential round trips prepare the system of
 coupled triggers in  thermal equilibrium state.
 For transition probabilities be less than unit the value of  $\ae$ should obey:

\begin{equation}
0\le\ae\le\exp\bigg(-\beta m |W|\bigg),
\end{equation}
here $m$ -- is the number of nearest neighbours. In our opinion it is possible
to create a specialized processor working in accordance with this algorithm.
The described algorithm can be implemented on conventional digital computer.
In this case $\ae$ should take the  maximum possible value
 $\ae=\exp\bigg(-\beta m |W|\bigg)$ for system to
relax as fast as possible.
 In relaxed system one can observe
 magnetisation $S=\sum_i s_i$, energy (13), heat capacity $c=\partial \langle H\rangle/\partial T$,
   an arbitrary correlation functions.

To demonstrate the aforesaid algorithm we simulate  the phase transition in two-dimensional Ising
magnetic by means of conventional computer. Fig.1  shows the temperature dependence of
heat capacity (top),
 magnetisation (middle) and
energy (bottom) calculated for the case of $W=-1$ (ferromagnetic). Calculations were performed for
 lattice with sizes $1000\times 1000$.
The procedure was as follows. At the beginning   the system was
prepared in the state with magnetisation close to its ultimate
value $S=N$ and with temperature much lower than the temperature of phase transition $T_c$.
After that the above algorithm  started with
gradually increasing temperature $1/\beta$.
  When temperature becomes close to $T_c$  the magnetisation vanishes  and heat
 capacity takes its maximum value. The value of $T_c$ obtained in our
 calculations is in agreement with the exact formula of Kramers and
 Wannier \cite{Kra}.

 The similar calculations can be performed for the case of
  {\it zero} initial magnetisation $S=0$. In this case total magnetisation is zero  for all temperatures.
   Fig.2 shows the spatial distribution of 2D
 magnetisation below $T_c$ (top picture,  the domains are clearly seen) and above $T_c$ (bottom).
  The heat capacity and energy temperature behaviour is similar to that in
  fig.1.

The algorithm described can be directly  generalized for the case of Ising system with an arbitrary
interspin interaction $W(r)$:
$$
H={1\over 2}\sum_{ik}W(i-k)s_is_k
$$
To do this one should use the effective magnetic field in the form
$$
h_i=\sum_r W(r-i) s_r
$$
instead equation (5).

Possibly the described algorithm may be useful for checking the gauge
theories of critical phenomena \cite{Cas,Cas1,Panero}. In this case  the duality of some gauge
models with respect to Ising system is exploited. The similar algorithms were described in
 \cite{Cas,Cas1,Panero}.

\section{ 1D- correlation functions.}

To check the above algorithm let us consider the exactly
solvable one-dimensional Ising magnetic with nearest neighbours
interaction.
 This problem was solved by Ising \cite{Ising} but above algorithm provide a simple way
 to obtain formulas (27) for correlation functions which are not very popular.
In the case of 1D Ising magnetic with nearest neighbours interaction the field acting on the $i$-th spin
can be calculated by formula (5) as:

 \begin{equation}
 h_i=W(s_{i-1}+s_{i+1})
 \end{equation}

Suppose the clock pulse acts on $i$-th spin (trigger). Let us calculate
the increment
$\Delta\langle s_i f\rangle$,  where  $f$ is an arbitrary function
of of all spin variables except $s_i$.
 Multiplying both parts of equation (11) by $s_i f$ and
summing over all states $\Psi$ we obtain:

\begin{equation}
\Delta \langle s_i f\rangle=\ae\bigg(
-\sum_\Psi \sigma(\Psi)  f s_i \exp [\beta W(s_{i+1}+s_{i-1})s_i] \bigg|_\Psi
\end{equation}
$$
+\sum_\Psi \sigma(\hat{O}_i\Psi) f s_i \exp [-\beta W(s_{i+1}+s_{i-1})s_i] \bigg|_\Psi
\bigg)
$$

Passing  from summation over $\Psi$
to summation over $\hat{O}_i\Psi$ in the second sum
($s_i$ should be replaced by $-s_i$), denoting
\begin{equation}
\beta W\equiv \theta,
\end{equation}
 and using the relation
\begin{equation}
\exp\alpha s=\hbox{ch}\hskip0.5mm\alpha+s \hskip1mm\hbox{sh}\hskip0.5mm\alpha,
 \hskip20mm s=\pm1,
\end{equation}
 we obtain:
\begin{equation}
\Delta\langle s_i f\rangle=-2\ae\langle  f s_i \exp [\beta W(s_{i+1}+s_{i-1})s_i] \rangle=
\end{equation}
$$
= -2\ae\bigg(
\hbox{ch}\hskip0.3mm^2\theta \langle s_i f\rangle+{1\over 2}\hbox{sh}\hskip0.3mm 2\theta(\langle s_{i+1} f\rangle+\langle s_{i-1} f\rangle)
+\hbox{sh}\hskip0.3mm^2\theta \langle s_{i+1} s_i s_{i-1} f\rangle
\bigg)
$$
 Hence in the equilibrium state:
 \begin{equation}
\hbox{ch}\hskip0.3mm^2\theta \langle s_i f\rangle+{1\over 2}\hbox{sh}\hskip0.3mm 2\theta(\langle s_{i+1} f\rangle+\langle s_{i-1} f\rangle)
+\hbox{sh}\hskip0.3mm^2\theta \langle s_{i+1} s_i s_{i-1} f\rangle=0
 \end{equation}

Now we use this relationship to calculate the equilibrium correlation function:

\begin{equation}
k_p\equiv \langle s_i s_{i+p}\rangle.
\end{equation}

This function
depends only on the difference $p$ of its indexes.
Let $f=s_{i+p}$  in equation (21). Then we have:
 \begin{equation}
 k_p\hskip0.5mm\hbox{ch}\hskip0.3mm^2\theta +{1\over 2}
 \hskip0.5mm(k_{p+1}+k_{p-1})\hskip0.5mm\hbox{sh}\hskip0.3mm 2\theta +\hbox{sh}\hskip0.3mm^2\theta
 \langle s_{i+1}s_i s_{i-1}s_{i+p}\rangle=0
 \end{equation}
To calculate the correlation function entering the last term let
$f=s_{i+1} s_{i-1} s_{i+p}$ in equation (21). We have:
  \begin{equation}
\hbox{ch}\hskip0.3mm^2\theta \langle s_{i+1}s_i s_{i-1}s_{i+p}\rangle+{1\over 2}
 \hskip0.5mm(k_{p+1}+k_{p-1})\hskip0.5mm\hbox{sh}\hskip0.3mm 2\theta
+k_p\hskip0.5mm\hbox{sh}\hskip0.3mm^2\theta=0
\end{equation}
Hence:
\begin{equation}
\langle s_{i+1}s_i s_{i-1}s_{i+p}\rangle=-k_p\hskip0.5mm \hbox{th}\hskip0.3mm^2\theta-
[k_{p+1}+k_{p-1}]\hskip0.5mm\hbox{th}\hskip0.3mm\theta.
\end{equation}

By substituting (25) in to (23) one can obtain:

\begin{equation}
k_p=\xi(k_{p+1}+k_{p-1}),\hskip20mm\xi\equiv -{1\over 2}\hbox{th}\hskip0.3mm(2\theta)
\end{equation}
The solution of this equation under condition $k_0=1$ has the form:
\begin{equation}
k_p=\exp(\alpha p),\hskip20mm \xi>0,\hbox{(ferromagnetic)}
\end{equation}
$$
k_p=(-1)^p \exp(\alpha p),\hskip20mm \xi<0,\hbox{(anti-ferromagnetic)}
$$

with $\alpha$  (for $p>0$) being the negative root of the equation:

\begin{equation}
\hbox{ch}\hskip0.3mm\alpha={1\over 2|\xi|}
\end{equation}

 Formulas (27) were verified by direct computer simulation according to
 the above algorithm for 1D Ising magnetic.

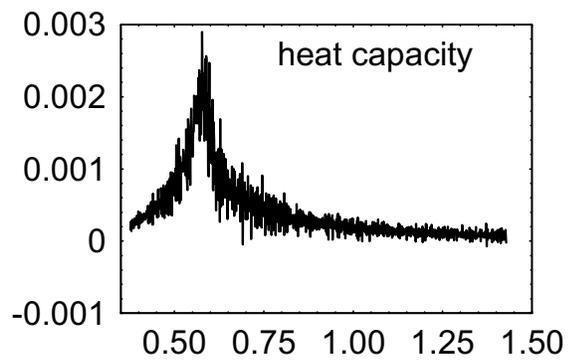
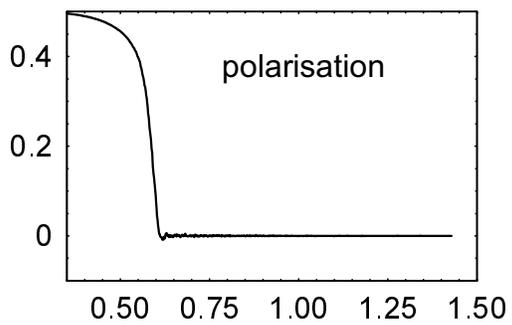
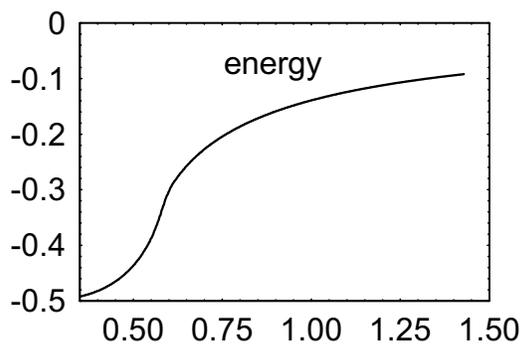
\end{document}